\documentclass[aps,preprint]{revtex4}%
\usepackage{amsfonts}
\usepackage{amsmath}
\usepackage{amssymb}
\usepackage{graphicx}%
\begin{document}
\title{Effect of the Born-Infeld Parameter in higher dimensional Hawking radiation }
\author{S. Habib Mazharimousavi$^{\ast}$}
\author{I. Sakalli$^{\ddag}$}
\author{M. Halilsoy$^{\dag}$}
\affiliation{Department of Physics, Eastern Mediterranean University,}
\affiliation{G. Magusa, north Cyprus, Mersin-10, Turkey}
\affiliation{$^{\ast}$habib.mazhari@emu.edu.tr}
\affiliation{$^{\dagger}$izzet.sakalli@emu.edu.tr}
\affiliation{$^{\ddag}$mustafa.halilsoy@emu.edu.tr}

\begin{abstract}
We show in detail that the Hawking temperature calculated from the surface
gravity is in agreement with the result of \ exact semi-classical radiation
spectrum for higher dimensional linear dilaton black holes in various
theories. We extend the method derived first by Cl\'{e}ment-Fabris-Marques for
$4-$dimensional linear dilaton black hole solutions to the higher dimensions
in theories such as Einstein-Maxwell-Dilaton, Einstein-Yang-Mills-Dilaton and
Einstein-Yang-Mills-Born-Infeld-Dilaton. Similar to the
Cl\'{e}ment-Fabris-Marques results, it is proved that whenever an analytic
solution is available to the massless scalar wave equation in the background
of higher dimensional massive linear dilaton black holes, an exact computation
of the radiation spectrum leads to the Hawking temperature $T_{H}$ in the high
frequency regime. The significance of the dimensionality on the value of
$T_{H}$ is shown, explicitly. For a chosen dimension, we demonstrate how
higher dimensional linear dilaton black holes interpolate between the black
hole solutions with Yang-Mills and electromagnetic fields by altering the
Born-Infeld parameter in aspect of measurable quantity $T_{H}$. Finally, we
explain the reason of, why massless higher dimensional linear dilaton black
holes cannot radiate.

\end{abstract}
\maketitle

\section{INTRODUCTION}

Although today there are several methods to compute the Hawking radiation,
(see for instance \cite{R1,R2,R3,R4,R5,R6}, and references therein), it still
attracts interest to consider alternative derivations. On the other hand, none
of them is completely conclusive. Nevertheless, the most direct is Hawking's
original study \cite{R1}, which computes the Bogoliubov coefficients between
in and out states for a realistic collapsing black hole. The most significant
remark on this study is that a black hole can emit particles from its event
horizon with a temperature proportional to its surface gravity. Another
elegant contribution was made to the Hawking radiation by Unruh \cite{R7}. He
showed that it is possible to obtain the same Hawking temperature $T_{H},$
when the collapse is replaced by appropriate boundary conditions on the
horizon of an eternal black hole. Instead of computing the Bogoliubov
coefficients in order to obtain the black hole radiation, one may
alternatively compute the reflection and transmission coefficients of an
incident wave by the black hole. This method works best if the wave equation
can be solved, exactly. From now on, we designate this method with
"semi-classical radiation spectrum method" and abbreviate it as \textit{SCRSM}.

Recently, Cl\'{e}ment, et al \cite{R8} have studied the \textit{SCRSM} for a
class of non-asymptotically flat charged massive linear dilaton black holes.
The metric of the associated linear dilaton black holes is a solution to the
Einstein-Maxwell-Dilaton (EMD) theory in 4-dimensions. It is shown that in the
high frequency regime, the \textit{SCRSM} for massive black holes yield the
same temperature with the surface gravity method. Their result for a massless
black hole is in agreement with the fact that a massless object cannot radiate.

In this Letter, we shall extend the application of \textit{SCRSM} to linear
dilaton black hole solutions in Einstein-Maxwell dilaton (EMD) \cite{R9},
Einstein-Yang-Mills dilaton (EYMD) \cite{R10} and
Einstein-Yang-Mills-Born-Infeld dilaton (EYMBID) \cite{R11} theories in higher
dimensions. The spacetimes describing these black holes are charged, dilatonic
and non-asymptotically flat. First, we introduce a generic line-element of
higher dimensional linear dilaton black holes in which the metric functions
are apt for the EMD, EYMD and EYMBID theories, where the latter two are
presented recently \cite{R10,R11}. Next, we consider the statistical $T_{H}$
of the massive linear dilaton black holes computed by using the surface
gravity and discuss their evaporation processes. According to the Stefan's
law, we show that higher dimensional linear dilaton black holes evaporate in
an infinite time. In the meantime, during the evaporation process, the Hawking
temperature remains constant for a given dimension. Besides this, the constant
value of $T_{H}$ increases with the dimensionality $N$. We then apply the
\textit{SCRSM} to the massive linear dilaton black holes and show that this
computation exactly matches with the statistical $T_{H}$ in the high frequency
regime. Finally, we answer the question, why the massless extreme black holes
do not radiate, by establishing a connection between our work and \cite{R8}.

The organization of the our Letter is as follows. In the next section, we
review briefly the higher dimensional linear dilaton black hole solutions in
the EMD, EYMD and EYMBID theories. In Sec.III, the evaporation of these black
holes are discussed according to the Stefan's law. Sec. IV is devoted to the
analytical computation of the $T_{H}$ via the \textit{SCRSM} for the massive
higher dimensional linear dilaton black holes. We plot some graphs to compare
the results acquired from each theory. We draw our conclusions in Sec. V.

\section{HIGHER DIMENSIONAL LINEAR DILATON BLACK HOLES IN EMD, EYMD AND EYMBID
THEORIES}

The metric ansatz for static spherically symmetric solutions representing
$N$-dimensional ($N\geq4$) linear dilaton black holes can be introduced by
\begin{equation}
ds^{2}=-fdt^{2}+\frac{dr^{2}}{f}+h^{2}d\Omega_{N-2}^{2},
\end{equation}
where $f$ and $h$ are only functions of $r$ and the spherical line element is
\begin{equation}
d\Omega_{N-2}^{2}=d\theta_{1}^{2}+\underset{i=2}{\overset{N-2}{%
%TCIMACRO{\tsum }%
%BeginExpansion
{\textstyle\sum}
%EndExpansion
}}\underset{j=1}{\overset{i-1}{%
%TCIMACRO{\tprod }%
%BeginExpansion
{\textstyle\prod}
%EndExpansion
}}\sin^{2}\theta_{j}\;d\theta_{i}^{2},
\end{equation}
in which $0\leq\theta_{k}\leq\pi$ with $k=1...N-3$, and $0\leq\theta_{N-2}%
\leq2\pi.$ Here, we set a proper ansatz for the metric functions $h$ as%

\begin{equation}
h=Ae^{-\frac{2\alpha\Phi}{N-2}},
\end{equation}
where $\Phi$ is the dilaton field, $\alpha$ is the dilaton parameter and $A$
is a coefficient to be determined for the respective theory. In the present
paper, dilaton parameter $\alpha$ for linear dilaton black holes is chosen by%

\begin{equation}
e^{-\frac{2\alpha\Phi}{N-2}}=\sqrt{r}\text{ \ }\rightarrow\text{\ }h=A\sqrt
{r},
\end{equation}
The field equations, which are obtained from the action of the theory together
with metric (1) suggest that the general form of the metric function $f$\ is%

\begin{equation}
f=\Sigma r\left[  1-(\frac{r_{+}}{r})^{\frac{N-2}{2}}\right]  ,
\end{equation}
where $\Sigma$ is another coefficient to be determined for each theory. From
now on, $r_{+}$ will be interpreted as the event horizon of the black hole. By
following the mass definition for the non-asymptotically flat black holes, the
so-called quasi-local mass $M$ introduced by Brown and York \cite{R12}, one
can see that the horizon $r_{+}$ is related to the mass $M$ and the dimension
$N$ through%

\begin{equation}
r_{+}=\left[  \frac{8M}{(N-2)\Sigma A^{N-2}}\right]  ^{\frac{2}{N-2}}.
\end{equation}

Higher dimensional linear dilaton black holes to the EMD theory was found long
time ago by Chan et al\textit{.} \cite{R9}. The solution is obtained from the
following $N$-dimensional EMD action%

\begin{equation}
I=\frac{-1}{16\pi}\int d^{N}x\sqrt{-g}\left(  R-\frac{4}{N-2}\left(
\nabla\Phi\right)  ^{2}-e^{-\frac{4\alpha\Phi}{N-2}}F^{2}\right)  ,
\end{equation}
where $F^{2}=F_{\mu\nu}F^{\mu\nu}$ for the Maxwell field. The coefficients $A$
and $\Sigma$ that give the correct metric functions (4) and (5) through the
action for the EMD theory are given by \cite{R9}%

\begin{equation}
\Sigma\rightarrow\Sigma_{EMD}=\frac{4}{\gamma^{2}}\left(  \frac{N-3}%
{N-2}\right)  ^{2}\text{ \ and }A\rightarrow A_{EMD}=\gamma,
\end{equation}
where $\gamma$ is a constant.

Besides the higher dimensional linear dilaton black hole solutions to the EMD
theory, new $N$-dimensional linear dilaton black hole solutions to the EYMD
and EYMBID theories are considered in the literature \cite{R10,R11}. The
actions are%

\begin{equation}
\mathbf{I}=-\frac{1}{16\pi}\int\nolimits_{\mathcal{M}}d^{N}x\sqrt{-g}\left[
R-\frac{4}{N-2}\left(  \nabla\Phi\right)  ^{2}+%
%TCIMACRO{\tciLaplace}%
%BeginExpansion
\mathcal{L}%
%EndExpansion
(\Phi)\right]  -\frac{1}{8\pi}\int\nolimits_{\partial\mathcal{M}}d^{N-1}%
x\sqrt{-h}K,
\end{equation}
and%

\begin{equation}
\mathbf{I}=-\frac{1}{16\pi}\int\nolimits_{\mathcal{M}}d^{N}x\sqrt{-g}\left[
R-\frac{4}{N-2}\left(  \nabla\Phi\right)  ^{2}+%
%TCIMACRO{\tciLaplace}%
%BeginExpansion
\mathcal{L}%
%EndExpansion
(\mathbf{F},\Phi)\right]  -\frac{1}{8\pi}\int\nolimits_{\partial\mathcal{M}%
}d^{N-1}x\sqrt{-h}K,
\end{equation}
which describe the EYMD and EYMBID theories, respectively. Here,%

\begin{equation}%
%TCIMACRO{\tciLaplace}%
%BeginExpansion
\mathcal{L}%
%EndExpansion
(\Phi)=e^{-\frac{4\alpha\Phi}{N-2}}\mathbf{Tr}(F_{\lambda\sigma}%
^{(a)}F^{(a)\lambda\sigma}),
\end{equation}

\begin{equation}%
%TCIMACRO{\tciLaplace}%
%BeginExpansion
\mathcal{L}%
%EndExpansion
(\mathbf{F},\Phi)=4\beta^{2}e^{\frac{4\alpha\Phi}{N-2}}\left(  1-\sqrt
{1+\frac{e^{-\frac{8\alpha\Phi}{N-2}}\mathbf{Tr}(F_{\lambda\sigma}%
^{(a)}F^{(a)\lambda\sigma})}{2\beta^{2}}}\right)  ,
\end{equation}
in which%

\begin{equation}
\mathbf{Tr}(.)=\overset{\left(  N-1\right)  (N-2)/2}{\underset{a=1}{%
%TCIMACRO{\tsum }%
%BeginExpansion
{\textstyle\sum}
%EndExpansion
}\left(  .\right)  .}%
\end{equation}

In the actions (9) and (10) $R$ is the usual curvature scalar, $\mathbf{F}%
^{\left(  a\right)  }=F_{\mu\nu}^{\left(  a\right)  }dx^{\mu}\wedge dx^{\nu}$
stands for the Yang-Mills (YM) 2-forms and $\beta$ denotes the Born-Infeld
parameter. The second term in the actions (9) and (10) is the surface integral
with its induced metric $h_{ij}$ and trace $K$ of its extrinsic curvature. It
is found that the corresponding coefficients to the metric functions (4) and
(5) of the $N$-dimensional linear dilaton black hole solutions to the EYMD
theory \cite{R10} are
\begin{equation}
\Sigma\rightarrow\Sigma_{EYMD}=\frac{\left(  N-3\right)  }{\left(  N-2\right)
Q^{2}}\text{ \ and \ }A\rightarrow A_{EYMD}=\sqrt{2}Q,
\end{equation}
and to the EYMBID theory \cite{R11} obtained as follows%

\begin{equation}
\Sigma\rightarrow\Sigma_{EYMBID}=\frac{2\left(  N-3\right)  }{\left(
N-2\right)  Q_{c}^{2}}\left[  1-\sqrt{1-\frac{Q_{c}^{2}}{Q^{2}}}\right]
\text{ \ and \ }A\rightarrow A_{EYMBID}=\sqrt{2}Q\left(  1-\frac{Q_{c}^{2}%
}{Q^{2}}\right)  ^{\frac{1}{4}}.
\end{equation}
Here $Q$ is known as YM charge and $Q_{c}$ is the critical value of YM charge
in which $Q^{2}>Q_{c}^{2}$ guarantees the existence of the metric in the
EYMBID theory. The value of the $Q_{c}^{2}$ is given as%

\begin{equation}
Q_{c}^{2}=\frac{(N-2)(N-3)}{8\beta^{2}}.
\end{equation}

\section{EVAPORATION OF HIGHER DIMENSIONAL LINEAR DILATON BLACK HOLES}

It can be seen from the metric function (5) that for $r_{+}>0$, the horizon at
$r=r_{+}$\ hides the null singularity at $r=0$. On the other hand, in the
extreme case $r_{+}=0$ metric (1) still exhibits the features of the black
holes. Since the central singularity $r=0$ is null and marginally trapped, it
prevents outgoing signals to reach external observers. Using the conventional
definition of the statistical Hawking temperature \cite{R13}, we get%
\begin{equation}
T_{H}=\frac{1}{4\pi}f^{\prime}\left(  r_{+}\right)  =\frac{\left(  N-2\right)
}{8\pi}\Sigma.
\end{equation}
One can immediately observe that $T_{H}$ is constant for an arbitrary
dimension $N$\ and increases with the dimensionality of the spacetime. As we
learned from the black body radiation, radiating objects loose mass in
accordance with the Stefan's law \cite{R8}. Therefore while a black hole
radiates, it should also loose from its mass. According to Stefan's law, we
should first calculate the surface area of the black hole (1). The horizon
area $S_{H}$ is found as%

\begin{equation}
S_{H}=\frac{2\pi^{\frac{N-1}{2}}}{\Gamma(\frac{N-1}{2})}h^{N-2},
\end{equation}
where $\Gamma(z)$ stands for the gamma function. After assuming that only
neutral quanta are radiated, Stefan's law admits the following time-dependent
horizon solutions%

\begin{equation}
r_{+}(t)=\bigskip\left\{
\begin{tabular}
[c]{lr}%
$\exp\left(  -\frac{1}{2\gamma^{6}}\left(  \frac{N-3}{N-2}\right)  ^{3}%
\mu(t)\right)  $ & EMD\\
$\exp\left(  -\frac{1}{2^{7}Q^{6}}\mu(t)\right)  $ & EYMD\\
$\exp\left(  -\frac{1}{2^{4}Q^{6}(1+\alpha)^{3}}\mu(t)\right)  $ & EYMBID
\end{tabular}
\ \ \right.
\end{equation}
where%

\begin{equation}
\mu\left(  t\right)  =\frac{\sigma\left(  N-3\right)  ^{3}\pi^{(\frac{N-9}%
{2})}}{\left(  N-2\right)  \Gamma(\frac{N-1}{2})}\left(  t-t_{0}\right)
\text{ \ \ and \ \ }0<\alpha=\sqrt{1-\frac{\left(  N-2\right)  \left(
N-3\right)  }{8\beta^{2}Q^{2}}}\leq1,
\end{equation}
in which $\sigma$\ is Stefan's constant, and $t_{0}$ is an integration
constant. From the results (19), we remark that $T_{H}$ is constant with
decreasing mass for a chosen dimension $N$, and the black holes reach to their
extreme states $r_{+}=0$ in an infinite time. Namely, the required time to
evaporate each black hole is infinite.

\section{CALCULATION OF $T_{H}$ VIA SCRSM}

Following the \textit{SCRSM }\cite{R8}, we now derive a more precise
expression for the temperature of the higher dimensional linear dilaton black
holes (1). To this end, we should first study the wave scattering in such
spacetimes (1) with Eqs. (4) and (5). Contrary to the several black hole
cases, here the massless wave equation
\begin{equation}
\square\Psi=0,
\end{equation}
admits an exact solution in the spacetimes (1). The Laplacian operator on a
$N$-dimensional metric is given by%
\begin{equation}
\square=\frac{1}{\sqrt{-g}}\partial_{\upsilon}\left(  \sqrt{-g}\partial
^{\upsilon}\right)  ,
\end{equation}
where $\upsilon$ runs from $1$\ to $N.$ One may consider a separable solution
as%
\begin{equation}
\Psi=R\left(  r\right)  e^{-i\omega t}Y_{l}\left(  \Omega_{N-2}\right)  ,
\end{equation}
in which $Y_{l}\left(  \Omega_{N-2}\right)  $ is the eigenfunction of $N-2$
dimensional Laplace-Beltrami operator $\nabla_{N-2}^{2}$ with the eigenvalue
$-l(l+N-3)$ \cite{R14}. After substituting harmonic eigenmodes (23) into the
wave equation (21) and making a straightforward calculation, one obtains the
radial equation:%
\begin{equation}
\partial_{r}\left[  h^{N-2}f\partial_{r}R\left(  r\right)  \right]
+h^{N-2}\left[  \frac{\omega^{2}}{f}-\frac{l\left(  l+N-3\right)  }{h^{2}%
}\right]  R(r)=0.
\end{equation}
After changing the independent variable and the parameters as
\begin{align}
y  &  =1-\left(  \frac{r}{r_{+}}\right)  ^{\left(  \frac{N-2}{2}\right)
},\text{ \ \ \ }\tilde{\lambda}^{2}=\frac{4}{\left(  N-2\right)  ^{2}\Sigma
A^{2}}l\left(  l+N-3\right)  ,\nonumber\\
\tilde{\omega}  &  =\varepsilon\omega,
\end{align}
where%

\begin{equation}
\varepsilon=\frac{2}{(N-2)\Sigma},
\end{equation}
one transforms the radial equation (24) into the following hypergeometric
equation
\begin{equation}
\partial_{y}\left[  y\left(  y-1\right)  \partial_{y}R\left(  y\right)
\right]  +\left(  \tilde{\omega}^{2}\frac{y-1}{y}-\tilde{\lambda}^{2}\right)
R\left(  y\right)  =0.
\end{equation}
Further, letting%

\begin{equation}
\tilde{\Lambda}=2ik,
\end{equation}
where $k$ is%

\begin{equation}
k=\sqrt{\tilde{\omega}^{2}-\tilde{\lambda}^{2}-\frac{1}{4}},
\end{equation}
(throughout the paper we assume that $k$ has a real value.), we can obtain the
general solution of (27) as follows%

\begin{align}
R\left(  y\right)   &  =C_{1}(-y)^{i\tilde{\omega}}F\left(  \frac{1}%
{2}+i(\tilde{\omega}+k),\frac{1}{2}+i(\tilde{\omega}-k),1+2i\tilde{\omega
};y\right)  +\nonumber\\
&  C_{2}(-y)^{-i\tilde{\omega}}F\left(  \frac{1}{2}+i(-\tilde{\omega}%
+k),\frac{1}{2}-i(\tilde{\omega}+k),1-2i\tilde{\omega};y\right)  .
\end{align}
Thus, the solution (30) leads to the general solution of Eq. (24) as%
\begin{align}
R\left(  \rho\right)   &  =C_{1}\left(  \frac{\rho-\tau}{\tau}\right)
^{i\tilde{\omega}}F\left(  \frac{1}{2}+i(\tilde{\omega}+k),\frac{1}%
{2}+i(\tilde{\omega}-k),1+2i\tilde{\omega};\frac{\tau-\rho}{\tau}\right)
+\nonumber\\
&  C_{2}\left(  \frac{\rho-\tau}{\tau}\right)  ^{-i\tilde{\omega}}F\left(
\frac{1}{2}+i(-\tilde{\omega}+k),\frac{1}{2}-i(\tilde{\omega}+k),1-2i\tilde
{\omega};\frac{\tau-\rho}{\tau}\right)  ,
\end{align}
in which
\begin{align}
\rho &  =\left(  r\right)  ^{\left(  \frac{N-2}{2}\right)  },\nonumber\\
\tau &  =\left(  r_{+}\right)  ^{\left(  \frac{N-2}{2}\right)  }.
\end{align}

Letting%

\begin{equation}
\frac{\rho-\tau}{\tau}=e^{x/\varepsilon},
\end{equation}
one gets the behavior of the partial wave near the horizon ($r\rightarrow
r_{+}$) as%

\begin{equation}
\Psi\simeq C_{1}e^{i\omega(x-t)}+C_{2}e^{-i\omega(x-t)}.
\end{equation}
where $C_{1}$ and $C_{2}$ are the amplitudes of the near-horizon outgoing and
ingoing waves.

Now, we shall use the one of the special features of the hypergeometric
functions in which it leads us to obtain the asymptotic behavior of the
partial wave. The feature is nothing but a transformation of the
hypergeometric functions of argument $y$ in (31) to the hypergeometric
functions of argument $1/y$. The relevant transformation is given by
\cite{R15}%

\begin{align}
F(a,b;c;y)  &  =\frac{\Gamma(c)\Gamma(b-a)}{\Gamma(b)\Gamma(c-a)}%
(-y)^{-a}F(a,a+1-c;a+1-b;1/y)\nonumber\\
&  +\frac{\Gamma(c)\Gamma(a-b)}{\Gamma(a)\Gamma(c-b)}(-y)^{-b}%
F(b,b+1-c;b+1-a;1/y).
\end{align}
This transformation yields the partial wave near spatial infinity as%

\begin{equation}
\Psi\simeq(\frac{r}{r_{+}})^{\frac{2-N}{4}}\left\{  B_{1}\exp\left[
i(\frac{k}{\varepsilon}x-\omega t)\right]  +B_{2}\exp\left[  -i(\frac
{k}{\varepsilon}x+\omega t)\right]  \right\}  ,
\end{equation}
where $B_{1}$ and $B_{2}$ denote the amplitudes of the asymptotic outgoing and
ingoing waves, respectively. After a straightforward calculation, one may
derive the relations between $B_{1}$,$B_{2}$ and $C_{1}$, $C_{2}$ as follows%

\[
B_{1}=C_{1}\frac{\Gamma(\widehat{c})\Gamma(\widehat{a}-\widehat{b})}%
{\Gamma(\widehat{a})\Gamma(\widehat{c}-\widehat{b})}+C_{2}\frac{\Gamma
(2-\widehat{c})\Gamma(\widehat{a}-\widehat{b})}{\Gamma(\widehat{a}-\widehat
{c}+1)\Gamma(1-\widehat{b})},
\]

\begin{equation}
B_{2}=C_{1}\frac{\Gamma(\widehat{c})\Gamma(\widehat{b}-\widehat{a})}%
{\Gamma(\widehat{b})\Gamma(\widehat{c}-\widehat{a})}+C_{2}\frac{\Gamma
(2-\widehat{c})\Gamma(\widehat{b}-\widehat{a})}{\Gamma(\widehat{b}-\widehat
{c}+1)\Gamma(1-\widehat{a})},
\end{equation}
where%

\begin{equation}
\widehat{a}=\frac{1}{2}+i(\tilde{\omega}+k),\text{ \ }\widehat{b}=\frac{1}%
{2}+i(\tilde{\omega}-k),\text{ \ }\widehat{c}=1+2i\tilde{\omega}.
\end{equation}

The coefficient of reflection by the black hole is calculated by virtue of the
fact that outgoing mode must be absent at the spatial infinity. This is
because the Hawking radiation is considered as the inverse scattering by the
black hole. Briefly $B_{1}=0$ and it naturally leads to%

\begin{equation}
R=\frac{\left\vert C_{1}\right\vert ^{2}}{\left\vert C_{2}\right\vert ^{2}%
}=\frac{\left\vert \Gamma(\widehat{a})^{2}\right\vert ^{2}}{\left\vert
\Gamma(\widehat{a}-\widehat{c}+1)^{2}\right\vert ^{2}},
\end{equation}
which is equivalent to%

\begin{equation}
R=\frac{\cosh^{2}\pi(k-\tilde{\omega})}{\cosh^{2}\pi(k+\tilde{\omega})}.
\end{equation}
Thus the resulting radiation spectrum is
\begin{equation}
\left(  e^{\frac{\omega}{T_{H}}}-1\right)  ^{-1}=\frac{R}{1-R}=\frac{\cosh
^{2}\pi\left(  k-\tilde{\omega}\right)  }{\cosh^{2}\pi\left(  k+\tilde{\omega
}\right)  -\cosh^{2}\pi\left(  k-\tilde{\omega}\right)  }.
\end{equation}

From here one may easily read the temperature%

\begin{equation}
T_{H}=\frac{\omega}{2\ln\left[  \frac{\cosh\pi\left(  k+\tilde{\omega}\right)
}{\cosh\pi\left(  k-\tilde{\omega}\right)  }\right]  },
\end{equation}
and for high frequencies $k\simeq\tilde{\omega}=\frac{2}{(N-2)\Sigma}\omega,$
Eq. (42) reduces to
\begin{align}
T_{H}  &  \simeq\underset{\omega\rightarrow\text{large value}}{\lim}%
\frac{\omega}{2\ln\left[  \frac{\cosh\pi\left(  k+\tilde{\omega}\right)
}{\cosh\pi\left(  k-\tilde{\omega}\right)  }\right]  }\simeq\frac{\omega}%
{2\ln(\cosh2\pi\tilde{\omega})},\nonumber\\
&  \simeq\frac{N-2}{8\pi}\Sigma.
\end{align}
which is nothing but the statistical Hawking temperature (17), which we
obtained before.

We plot $T_{H}$ (42) versus frequency $\omega$ for each theory with $N=5$, and
display all graphs in Fig. (1). As it can be seen from the Fig. (1), in the
high limits of the Born-Infeld parameter $\beta,$ the thermal behavior of the
linear dilaton black holes in the EYMBID theory exhibits similar behavior to
the EYMD theory. For a particular choice of $\beta$, it is possible to see the
common behaviors in thermal manner for the linear dilaton black holes in the
EYMBID and EMD theories. So we can deduce that in a special range of the
Born-Infeld parameter $\beta$, the linear dilaton black holes in the EYMBID
theory interpolate thermally between the black holes in the EYMD and EMD
theories. However, for $\omega\rightarrow\infty,$ $T_{H}$ reduces to the
almost same constant value for each theory. The next figure, Fig. (2) is to
examine $T_{H}$ versus dimension $N$ within the high frequency regime.
According to Eq. (43), Fig. (2) represents $T_{H}$ increasing linearly\ with
$N$ for the linear dilaton black holes in the EMD and EYMD theories, it
increases parabolically in the EYMBID case. On the other hand, the similar
behavior, where at the high limits of the Born-Infeld parameter $\beta$ the
behavior of the $T_{H}$ in the EYMBID theory is almost close to the behavior
of $T_{H}$ in the EYMD theory, is also observed in Fig. (2) as highlighted in
Fig. (1). On the other hand, in the EYMBID theory $T_{H}$ takes limited real
values depending on the Born-Infeld parameter $\beta$\ through the dimension
$N$. In the EYMBID theory, $T_{H}$ is real as long as the dimension $N$
satisfies the condition%

\begin{equation}
\frac{(N-2)(N-3)}{8Q^{2}}<\beta.
\end{equation}

If one studies the case $r_{+}=0$ (i.e. the case of extreme massless black
holes), the above analysis for computing the Hawking radiation fails. In
\cite{R8}, it is successfully shown that the wave scattering problem in the
extreme four dimensional linear dilaton black holes in the EMD theory reduces
to the propagation of eigenmodes of a free Klein-Gordon field in
two-dimensional Minkowski spacetime with an effective mass. Conclusively,
there is no reflection, so that the extreme linear dilaton black holes cannot
radiate, although their surface gravities remain finite. Since setting
$r_{+}=0$ reduces metric (1) to a conformal product $M_{2}\times S^{N-2}$ of a
two dimensional Minkowski spacetime with the $\left(  N-2\right)  $-sphere of
constant radius, the same interpretation is valid also for the extreme higher
dimensional linear dilaton black holes in the EMD, EYMD and EYMBID theories.
In summary, the massless higher dimensional linear dilaton black holes in the
EMD, EYMD and EYMBID theories cannot radiate as well.

\section{Conclusion}

In this paper, we have effectively utilized the \textit{SCRSM} to derive the
Hawking temperature $T_{H}$ for massive, higher-dimensional $(N\geq4),$ linear
dilatonic black holes in the EMD, EYMD and EYMBID theories. To do this, first
we have attempted to solve the massless scalar wave equation, exactly. Exact
solution of the wave equation plays a crucial role in deriving a more precise
result of the temperature of those non-asymptotically flat black holes. After
finding the solution in terms of the hypergeometric functions and using their
intriguing features, we have demonstrated that in the high frequency regime,
the results of \textit{SCRSM} agree with the temperature obtained from the
surface gravity for all considered theories.

One of the main results obtained from the Stefan's law is that as in the case
$N=4$ \cite{R8} the linear dilaton black holes evaporate in an infinite time,
for $N\geq5$ as well. The figures of $T_{H}$ have some important results which
are summarized as follows: (i) When the dimension $N$ is fixed, the behavior
of $T_{H}$ versus frequency $\omega$ in the EYMBID theory exhibits similar
behavior of the $T_{H}$ in the EYMD theory with large $\beta$. (ii) From the
thermal point of view, for a special range of $\beta$ the linear dilaton black
hole solutions to the EYMBID theory interpolate between the black hole
solutions to the EMD and EYMD theories. (iii) Contrary to the EMD and EYMD
theories, in the EYMBID theory, at high frequency regime, $T_{H}$ increases
with $N$ parabolically rather than linearly. (iv) In the EYMBID theory,
$T_{H}$ is real unless the condition $\frac{(N-2)(N-3)}{8Q^{2}}<\beta$ is violated.

We also verify that contrary\ to the non-zero values of their surface gravity
the massless, extreme higher dimensional linear dilaton black holes\ do not
radiate. Finally, we remark that since our dilatonic black holes are
conformally related to the Brans-Dicke black holes \cite{R16} our results can
be extended to the latter theory as well.

\bigskip

\section{Figure Captions}

Figure 1:Hawking temperature $T_{H\text{ }}$ as a function of $\omega$ in 5D.
The relation is given by (42). Different line styles belong to different
theories: Dotted lines represent the EMD, dashed lines represent the EYMD and
solid lines correspond to the EYMBID. The physical parameters in (42) are
chosen as follows: $l=1$, $Q=1$ and $\gamma=\sqrt{2}.$

Figure 2: A plot of the high frequency limits of Hawking temperatures
$T_{H\text{ }}$ versus the dimension number $N$\ of the spacetime (1). Eq.
(43) or Eq. (17) governs the plots. Different line styles belong to different
theories: Dotted lines represent the EMD, dashed lines represent the EYMD and
solid lines correspond to the EYMBID. The physical parameters in (43) are
chosen as follows: $l=1$, $Q=1$ and $\gamma=\sqrt{2}.$


\begin{thebibliography}{99}                                                                                               %
\bibitem {R1}S. W. Hawking, Nature (London) \textbf{248}, 30 (1974); S. W.
Hawking, Commun. Math. Phys. \textbf{43}, 199 (1975).

\bibitem {R2}T. Damour and R. Ruffini, Phys. Rev. D \textbf{14,} 332 (1976);
S.Q. Wu and X. Cai, J. Math. Phys. \textbf{44}, 1084 (2003).

\bibitem {R3}M. K. Parikh and F. Wilczek, Phys. Rev. Lett. \textbf{85}, 5042
(2000); S. P. Robinson and F. Wilczek, Phys. Rev. Lett. \textbf{95}, 011303 (2005).

\bibitem {R4}S. Iso, H. Umetsu, and F. Wilczek, Phys. Rev. Lett. \textbf{96},
151302 (2006); S. Iso, H. Umetsu, and F. Wilczek, Phys. Rev. D. \textbf{74},
044017 (2006); K.Murata and J. Soda, Phys. Rev. D \textbf{74}, 044018 (2006).

\bibitem {R5}Q. Q. Jiang and S. Q. Wu, Phys. Lett. B \textbf{635}, 151 (2006);
Q. Q. Jiang, S. Q. Wu, and X. Cai, Phys. Rev. D \textbf{73}, 064003 (2006); Q.
Q. Jiang, S. Q. Wu, and X. Cai, Phys. Rev. D \textbf{75}, 064029 (2007); Q. Q.
Jiang, S. Q. Wu, and X. Cai, Phys. Lett. B \textbf{651}, 58 (2007); Q. Q.
Jiang, Class. Quantum Grav. \textbf{24}, 4391 (2007).

\bibitem {R6}R. Banerjee and S. Kulkarni, Phys. Rev. D \textbf{77}, 024018
(2008); R. Banerjee and S. Kulkarni, Phys. Lett. B \textbf{659}, 827 (2008).

\bibitem {R7}W.G. Unruh, Phys. Rev. D \textbf{14}, 870 (1976).

\bibitem {R8}G. Cl\'{e}ment, J.C. Fabris, and G.T. Marques, Phys. Lett. B
\textbf{651,} 54 (2007).

\bibitem {R9}K. C. K. Chan, J. H. Horne and R. B. Mann, Nucl. Phys. B
\textbf{447,} 441 (1995).

\bibitem {R10}S. H. Mazharimousavi, M. Halilsoy, Phys. Lett. B \textbf{659,}
471 (2008).

\bibitem {R11}S. H. Mazharimousavi, M. Halilsoy and Z. Amirabi, "N-Dimensional
non-abelian dilatonic, stable black holes and their Born-Infeld extension "arXiv:0802.3990.

\bibitem {R12}J. D. Brown and J.W. York, Phys. Rev. D \textbf{47} 1407 (1993).

\bibitem {R13}R.M. Wald, \textit{General Relativity} (The University of
Chicago Press, Chicago and London, 1984).

\bibitem {R14}D. P. Du and B. Wang, Phys. Rev. D \textbf{70}, 064024 (2004).

\bibitem {R15}M. Abramowitz and I. A. Stegun, \textit{Handbook of Mathematical
Functions} (Dover, New York, 1965).

\bibitem {R16}R. G. Cai and Y. S. Myung, Phys. Rev. D \textbf{56} (1997) 3466.
\end{thebibliography}
\end{document}